\documentclass[prb,twocolumn,showpacs]{revtex4}

\begin{document}\relax

\preprint{RU/4-03g}

\title{Note on nonequilibrium stationary states and entropy}

\author{G. Gallavotti, E.G.D. Cohen}
\affiliation{
Rutgers Hill Center, I.N.F.N. Roma1, Fisica Roma1 and
Rockefeller University}

\date{11 december 2003}
\begin{abstract}
In transformations between nonequilibrium stationary states, entropy
might be a not well defined concept. It might be analogous to the
``heat content'' in transformations in equilibrium which is not well
defined either, if they are not isochoric ({\it i.e.} do involve
mechanical work). Hence we conjecture that un a nonequilbrium
stationary state the entropy is just a quantity that can be
transferred or created, like heat in equilibrium, but has no physical
meaning as ``entropy content'' as a property of the
system.\end{abstract}


\maketitle

\centerline{\bf I. Thermostats and chaotic hypothesis.}
\vskip3mm 
In studying equilibrium and nonequilibrium thermodynamics the notion
of {\it thermostat} plays an important role: it is usually defined
phenomenologically\cite{Ze68} as a system capable only of
exchanging heat without changing its temperature or performing work (hence
it is ideally an infinite system).

In a statistical mechanical approach to a theory of the nonequilibrium
stationary states of a system which is subject to external non
conservative forces, which (therefore) perform work, thermostats must
then be present to avoid that the work performed by the external
forces causes an increase of the energy beyond bounds. In statistical
approaches the thermostats must be defined as mechanical forces: the
routes that one can follow are \vskip3mm 

\noindent (i) Introduce stochastic forces acting on the system (usually on the
boundary) \\ (ii) Assume that the system interacts via conservative
forces with infinite systems (``thermostats'') which are initially in
equilibrium and which one would like to show that by interacting with
the system the thermostats will undergo only changes localized in the
vicinity of the contact surfaces\\ (iii) Assume that the interaction
with the ``outside world'' is modeled by an effective force on the
system which balances the work of the external forces working on the
system ({\it cf.} Eq. (\ref{1e}) below) thus allowing the system to reach
stationarity.  \vskip3mm 

The third possibility has recently emerged as a very convenient way of
studying the problem, since it at least avoids the virtually
untractable theory of the behavior of systems in contact with infinite
reservoirs. However it is often regarded as ``unphysical'' because it
``amounts to modifying the system's equations of motion''. But the
equations of motion are modified also if one uses the option (i),
while the option (ii) is interesting but not really suitable for
investigations which, at least until now, mostly rely on numerical
experiments. Furthermore theoretically the only ``infinite
thermostats'' which have been considered are infinite systems of free
particles interacting only with the particles of the system and not
directly with each other, a rather restricted case.

Here we shall model thermostats by mechanical forces acting on the
system, however we do not want to make very specific choices of the
forces and the thermostats since we are interested only in general
properties which would be shared by ``any'' (reasonable) choices of
the forces and thermostat models. We think that the mechanical models
of thermostats fall into ``equivalence classes'', just as one thinks
that phenomenological thermostats do. Therefore we consider
mechanical systems which, in spite of being acted upon by
nonconservative forces, are kept in a nonequilibrium
stationary state with the help
of other mechanical (``thermostatic'') forces and study {\it which
relations, if any, can be established about transitions between
stationary states}\cite{Ga03}. We restrict ourselves here to
transformations between stationary states which are ``{\it quasi
static}'' transformations through intermediate stationary states. This
means that on the time scale of the observations the control
parameters ({\it e.g.} volume, strength of the external forces {\it etc.}) of the
stationary states are kept fixed long enough so that the system can at
any time can be considered to be in a stationary state. This is a
generalization of reversible transitions in equilibrium.

We consider only systems consisting of many particles and we do {\it
not} consider systems that are modeled by continua. Continua could be
considered but one first must understand the thermodynamics of simple
systems\cite{Ga02}.  The particle motions occurring in a simple system
are assumed to be chaotic: the {\it Chaotic Hypothesis}\cite{GC95},
essentially states that an isolated system of particles has a chaotic
evolution on microscopic time scales (see below).

A simple system will be described by a differential equation in its
phase space: we write it as
$\dot x=X_E(x)$ where $x=(\dot{\underline q},\underline q)\in R^{6N}
\equiv \Omega$ ({\it phase space}), $N$=number of particles, $m$= mass of
the particles, with

\begin{eqnarray}
m\ddot{\underline q}=f(\underline q)+\underline E\cdot \underline g(\underline
q)-\vartheta_{\underline E}
(\dot {\underline q},\underline q)\equiv X_E(x)\label{1e}
\end{eqnarray}
where $f(\underline q)$ describes the internal (conservative) forces
({\it e.g.} hard cores), $\underline E\cdot\underline g(\underline q)$
represents the ``active external forces'' (nonconservative for the
reasons mentioned at the beginning) acting on the system: for
definiteness we suppose that they are locally conservative (like an
electromotive force) but not globally, and
$\underline\vartheta_{\underline E}$ is the force law which models the
action exerted by the thermostat on the system to keep it from
indefinitely acquiring energy: this is why we shall call it a {\it
mechanical thermostat}. Linearity of the dependence on the ``fields''
$\underline E$ is only for convenience: we are {\it not} assuming them
to be small (the theory of linear nonequilibrium is amply discussed in
the literature\cite{DGM84}).  More generally the external forces could
be velocity dependent and even time dependent (periodically) but we
restrict ourselves to positional forces only for simplicity.

Models of thermostats in the above sense can be very different even
for the same macroscopic system; for instance (a list far from exhaustive)
\vskip3mm 

(1) assuming the system to have hard cores ({\it e.g.} a granular
material of the type considered in\cite{FN03}) one can consider
inelastic collisions, {\it e.g.} suppose that the head-on component of
the energy is decreased by a scale factor $\eta<1$ upon each collision
or, {\it alternatively}, the total energy of the two colliding
particles is rescaled after a collision and assigned a value
$2\frac32 k_B T$ (this is essentially ``Drude's conduction
model'')\cite{Be64}, or

(2) assuming that there is a background friction $\vartheta_i=-\nu \dot
    q_i$, $\nu>0$, for all components of $\ddot x_j$ or

(3) assuming {\it minimum effort} to keep, say, the total kinetic
    energy or the total energy constant, (``{\it Gaussian
    thermostat}''\cite{EM90}).
\vskip3mm 

{\it Remark:} the restriction that the external forces ($\underline
E\cdot\underline
g(\underline q)$) be positional is a strong restriction as it does not allow
velocity dependent forces or several different thermostats as needed
in any heat conduction problem where the nonequilibrium stationary
state is achieved by putting the system in contact with two reservoirs
without any external forces acting. However, such cases could also be
considered\cite{Ga96a}, but we do not treat them here since we want
to restrict ourselves to the simplest case.\vskip3mm 

We shall assume about the system which we consider the\vskip3mm 

{\it Chaotic Hypothesis\cite{GC95}: The
system evolution is supposed to be as chaotic as possible, {\it i.e.} to be
hyperbolic (one also says, technically, that the system is ``an Anosov
system'').}  \vskip3mm 

\noindent which will be a key assumption in our analysis.

\vskip3mm 
\centerline{\bf II. SRB statistics and nonequilibrium ensembles} \vskip3mm 
Any initial state $x$, randomly chosen in phase space with a
probability distribution which has a density in phase space ({\it i.e.} such
that the probability of a phase point $\dot {\underline q},\underline q$ to be
in
$d\dot {\underline q}d\underline q$ has the form $\rho(\dot {\underline
q},\underline q)d\dot {\underline q}d\underline
q$ for some probability density $\rho(\dot {\underline q},\underline q)\ge0$ in
phase
space), {\it will admit a statistics} (under the above chaotic
assumption): {\it i.e.} for all (smooth) observables $F$

\begin{eqnarray}
\lim_{T\to\infty} \frac1T\int_0^T F(S_tx) dt=\int_\Omega \mu_E(dy)
F(y)\label{2e}
\end{eqnarray}
where $\mu_{\underline E}$ is a stationary probability distribution on phase
space, called the {\it SRB distribution} or also {\it SRB
statistics}\cite{Ru95,Ga00,Ga02,GBG04}. This is a (nontrivial)
consequence of the above chaotic hypothesis.
\vskip3mm 

{\it Definition: A system in a microscopic state $x$, which has SRB
  statistics $\mu_{\underline E}$, is said to be in the stationary state
$\mu_{\underline
  E}$. The collection of all stationary states of a system that are
  constructed by varying the parameters (typically the volume $V$ of
  the container, the number of particles $N$, the external forces $\underline
E$,
  {\it etc}) will be called a ``nonequilibrium Ensemble''.}
\vskip3mm 

The {\it Ensemble} (with capital $E$) is therefore a collection of
probability distributions which we distinguish from what, following
Gibbs, has become established terminology where {\it ensemble}
indicates {\it a single element} of the collection, with fixed control
parameters: what is usually called simply the
``microcanonical ensemble with parameters $U,V$'' is here just a
single element of the collection ({\it i.e.} the Ensemble) of microcanonical
probability distributions.

The notion of Ensemble in non\-equilibrium is wider than in
equilibrium since it depends {\it also} on the equations of motion,
because of the presence there of the thermostats. However, one expects
that, as happens in equilibrium statistical mechanics, there should be
``equivalent Ensembles'' corresponding to classes of different
possible models for thermostats acting on a system\cite{Ga00,Ga02}.

Equilibrium is a special case of a nonequilibrium stationary state: in
such case $\underline E=\underline 0$ and $\underline\vartheta_{\underline
E}=\underline0$ and the chaotic
hypothesis implies the validity of the ergodic hypothesis\cite{Ga00};
the Ensemble (or collection) of SRB distributions each of which can be
parameterized by the total energy $U$ and volume $V$ {\it coincides}
with the corresponding collection of microcanonical
ensembles\cite{Ga00}. Furthermore, in general, the chaotic hypothesis
implies that observables that are represented by smooth functions on
phase space have finite time correlations which converge exponentially
fast to their stationary state averages ({\it i.e.} SRB averages).

We now want to consider which relations can be established in general
between the properties of stationary states that can be transformed into
one another by changing reversibly the external parameters, just as is done on
equilibrium states.

In fact, if we limit ourselves to equilibrium states first then it is well
known since Boltzmann (in his papers in the period 1866--1884,
see\cite{Bo84}) that if a transformation generates an energy variation
$dU$ and a volume variation $dV$ when the pressure (defined as a time
average of a function defined on phase space, see for
instance\cite{Ga00}) is $p$ and the average kinetic energy is $\frac32
N k_BT$ then, see\cite{Ga00} appendices A1.1 and A9.3,

\begin{eqnarray}
\frac{dU+p\, dV}{T}={\rm exact\ differential}
\label{3e}
\end{eqnarray}
while $dU + p\, dV$ is not exact, {\it except} in the isochoric case
({\it i.e.} when $dV=0$) and it is called the {\it heat transferred} from the
heat reservoirs to the system. It makes no sense to talk of {\it heat
content} contained in the system\cite{LL}, unless one limits oneself
to isochoric transformations: there is no heat content of a system
because one cannot distinguish between the heat and the mechanical
work contents unless one allows only isochoric transformations in
which the system performs no work (and in that case it is just another
name for the internal energy).

Defining the {\it entropy} content of a system as an integral $S$ of the exact
differential $(dU+p\, dV)/T$, an immediate question is whether one
can extend the notion of entropy content to non equilibrium
states.
\vskip3mm 

\centerline{\bf III. Entropy production rate and temperature.}
\vskip3mm 

The proposal that emerges from various theore\-ti\-cal considerations and
a number of numerical ex\-pe\-riments\cite{An82,EM90,Ru99,Ga00}, is to
define, if $k_B$ is Boltzmann's constant,
\vskip3mm 

{\it Definition: The entropy production rate in a stationary state
  $\mu_{\underline E}$ is $k_B\sigma_+$ with ({\it cf.} Eq.(\ref{2e})):

\begin{eqnarray}\sigma_+=
\langle{\sigma}\rangle\,{\buildrel def \over =}\,\int_\Omega \mu_{\underline
E}(dx)\sigma(x)\label{4e}\end{eqnarray}
where $\sigma(x)=-$ divergence of $\underline X_{\underline E}(x)$ ({\it i.e.}
$\sigma(x)$ is the
phase space contraction rate) and $\mu_{\underline E}$ is the SRB statistics.}
\vskip3mm 

This definition elicits a few comments: \vskip3mm 

\noindent (a) There is  no generally  accepted
definition of entropy in non\-equi\-librium stationary states.\\ (b)
In several thermostat models considered in the literature the average
divergence $\sigma_+$ of the equations of motion is essentially
related to $W$, the average work per unit time done by the
thermostatting forces, {\it i.e.} to the time average $W$ of $\dot {\underline
q}\cdot\underline\vartheta(\underline q,\dot{\underline q})$, which in
stationary states equals the
average work done by the external forces. For instance if
$\underline\vartheta(\underline
q,\dot{\underline q})$ is proportional to $\dot{\underline q}$, {\it i.e.}
$\underline\vartheta(\underline
q,\dot{\underline q})=\alpha(\dot{\underline q},\underline q)\,\dot {\underline
q}$ for some function $\alpha$,
then $W=\langle{\alpha \dot{\underline q}^2}\rangle$ while
$\sigma_+=\langle{\sigma}\rangle=3N\langle{\alpha}\rangle+
\langle{ (\partial_{\dot{\underline
q}}\,\alpha) \dot{\underline q}}\rangle$:
hence $W\simeq \langle{\alpha}\rangle 2K$ and 
$\sigma_+\simeq3N\langle{\alpha}\rangle$
so that in such cases $W$ and $\sigma_+$ are related by
$\sigma_+=\frac{W}{2K/3N}$. Since the work done by the thermostatting
forces is naturally interpreted as the heat that the system cedes to
the thermostat we see that in the cases considered ({\it i.e.}
$\underline\vartheta(\underline
q,\dot{\underline q})=\alpha(\dot{\underline q},\underline q)\,\dot {\underline
q}$) the quantity $\sigma_+ $
has the meaning of the entropy increase of the thermostat.  \\ (c) An
important general theorem\cite{Ru96}, guarantees that $\sigma_+\ge0$,
and $\sigma_+=0$ corresponds to the case in which the SRB distribution
$\mu_{\underline E}$ admits a density on phase space, a case that one naturally
identifies with an equilibrium state and which essentially happens
only if $\underline E=\underline0$.  \vskip3mm 

Certainly the above three properties are, at best, only an indication
that the phase space contraction can be interpreted as the entropy
increase of the thermostats (in the classical sense of the word and
due to the heat generated by the system). If we used phenomenological
thermostats they would be systems in thermal equilibrium and at a
fixed temperature so that the heat absorbed per unit time would
generate an entropy increase of the thermostats which is well defined.
Here one has to bear in mind that the the notion of heat absorbed by a
mechanical thermostat as well as the notion of its temperature are
{\it new concepts}. We use the arbitrariness offered us by the lack of a
generally accepted definition of these concepts to conjecture the
above definition of the entropy production rate on the basis of the
general result in (c) which guarantees the positivity that is desired
for compatibility with classical thermodynamics.

The notion of temperature of a thermostat is however still missing but
the above definition leads to a definition of an effective temperature
of the thermostatting forces (we stress that there is no universally
accepted definition of temperature in systems out of equilibrium, even
if stationary\cite{FN03,Ga03}). Here we propose
\vskip3mm 

{\it Definition: the (effective) temperature $T$ of the thermostats for a
stationary nonequilibrium state is

\begin{eqnarray} T=\frac{W}{k_B \sigma_+}\label{5e}\end{eqnarray}
where $W$ is the average work per unit time done on the system by the external
forces, equal to the average work $\dot Q$ done on the thermostatting forces,
and $k_B \sigma_+$ is the entropy production rate.}
\vskip3mm 

The equality between $W=\langle{\underline E\cdot\underline
g(\underline q)\cdot\dot{\underline q}}\rangle$ and $\dot
Q=\langle{\underline \vartheta\cdot\dot{\underline q}}\rangle$ is due
to the fact that the internal forces being conservative perform zero
work in the average.

{\it Remark:} the situations in which there is heat conduction between
different thermostats is not considered here: in such cases one has at
least two thermostats acting on the system: {\it i.e.} the
thermostatting force $\underline\vartheta$ is then the sum of, for
instance, forces
$\underline\vartheta=\underline\vartheta_1+\underline\vartheta_2$,
which perform the work $-\dot Q_1$ and $-\dot Q_2$, respectively, so
that the divergence $\sigma(x)$ is the sum of two quantities
$\sigma_1(x)$ and $\sigma_2(x)$. Therefore in such cases it will be
natural to define the temperatures of the two thermostats as
$T_i=\frac{\dot Q_i}{k_B \sigma_{i+}}$: we do not discuss the matter
further since, from the outset, we are not considering situations in
which the system is subject to several thermostatting forces.
\vskip3mm

The above definition does not make sense as such in equilibrium
because it becomes $0/0$: however, one can imagine introducing a small
forcing ${\underline  E}$ and a corresponding thermostat. Then in the limit of
vanishing forcing this yields a definition of $T$ which by the
``fluctuation dissipation theorem'' can be checked to be the correct
equilibrium temperature\cite{CKP97,Ga96a,Ga96,GR97}.

Our definition of nonequilibrium temperature has already been hinted
at by F. Bonetto and N. Menon as well as used in the
literature\cite{CKP97,FN03}, in particular cases.

Adopting the above concepts leads naturally to giving up the
possibility of defining the entropy content of a nonequilibrium
stationary state because the system creates entropy at a constant rate
and if one would insist in defining the entropy content of a
dissipating ({\it i.e.} with $\sigma_+>0$) stationary state one would
be compelled to assign to it a value $-\infty$. Thus in our view of
nonequilibrium stationary states the entropy ends up to be undefined and
one can speak meaningfully only of ``entropy production'' or ``entropy
transfer'': much as the ``heat content'' of a system is undefined in
equilibrium, but production and transfer of heat are well defined.
\vskip3mm 

\centerline{\bf IV. Discussion.}
\vskip3mm

(1) Having defined the notion of entropy production rate one can
define a ``duality'' between fluxes $J_j$ and forces $E_j$ using
$k_B\sigma_+=k_B\int \mu(dx)\sigma(x)$ as a ``generating function'':
$$J_j(\underline E)\,=\,k_B\,\frac{\partial \sigma_+}{\partial{E_j}}$$
which, in the limit $\underline E=\underline0$, leads to Onsager's reciprocity
and to
Green--Kubo's formulae for transport\cite{Ga96,GR97}.
\vskip3mm 

(2) We have proposed a general definition of entropy production rate
and of temperature for a class of stationary states. But a new
definition is really useful if it is associated with new results: we
think that such new results may already be around and cluster around
the {\it fluctuation theorem}, for which we refer to the
literature\cite{ECM93,GC95,BGG97,Ge98,GRS02,FN03,Ga03,CHGLPR03}.  \vskip3mm 

(3) The reason for our conjecture on the absence of an entropy content
    in nonequilibrium stationary states differs from the absence of
    heat content in equilibrium.  This because in equilibrium the heat
    content cannot be defined separately from the mechanical energy
    content, \cite{LL}. However in a nonequilibrium stationary state
    the impossibility to define entropy content is due to the steady
    entropy production, which makes the entropy content $-\infty$. In
    spite of that there is an analogy in that both quantities can be
    transferred or produced and they can even be defined if one limits
    oneself to consider a suitably restricted class of transformations
    ({\it e.g.} isochoric transformations between equilibrium states for what
    concerns heat or general transformations between equilibrium
    states for what concerns entropy).
\vskip3mm 

\noindent {\bf Acknowledgements:} {\it One of us (GG) is indebted to
stimulating discussions with F.Bonet\-to, S.Goldstein, J.Lebowitz,
O. Costin, D. Ruelle, E. Speer. EGDC gratefully acknowledges
financial support from the Office of Basic Engineering Sciences
of the US department of Energy under grant DE-FG02-88-ER13847.}

\kern3mm \noindent e-mail\\ {\tt
egdc@mail.rockefeller.edu,\\ gallavotti@roma1.infn.it}


\bibliographystyle{unsrt}
\vskip3mm
{R\kern-1mm\lower0.5mm\hbox{E}\kern-0.6mm V\kern-0.5mm%
\lower0.5mm\hbox{T}\kern-0.5mm E\kern-.5mm \lower0.5mm\hbox{X}}
\end{document}